\documentclass[sigconf]{acmart}

\AtBeginDocument{%
  \providecommand\BibTeX{{%
    \normalfont B\kern-0.5em{\scshape i\kern-0.25em b}\kern-0.8em\TeX}}}

\setcopyright{none}

\renewcommand\footnotetextcopyrightpermission[1]{}

\begin{document}

\title{Migration in the \textit{Stencil} Pluralist Cloud Architecture}

\author{Tai Liu}
\affiliation{%
  \institution{Tencent America LLC}
}
\email{tyeliu@tencent.com}

\author{Zain Tariq}
\affiliation{%
  \institution{New York University Abu Dhabi}
}
\email{zain.tariq@nyu.edu}

\author{Barath Raghavan}
\affiliation{%
  \institution{University of Southern California}
}
\email{barath.raghavan@usc.edu}

\author{Jay Chen}
\affiliation{
 \institution{International Computer Science Institute}
 }
 \email{jchen@icsi.berkeley.edu}

\begin{abstract}
A debate in the research community has buzzed in the background for years: should large-scale Internet services be centralized or decentralized?  Now-common centralized cloud and web services have downsides---user lock-in and loss of privacy and data control---that are increasingly apparent.  However, their decentralized counterparts have struggled to gain adoption, suffer from their own problems of scalability and trust, and eventually may result in the exact same lock-in they intended to prevent.

In this paper, we explore the design of a pluralist cloud architecture, \textbf{Stencil}, one that can serve as a narrow waist for user-facing services such as social media.  We aim to enable pluralism via a unifying set of abstractions that support migration from one service to a competing service. We find that migrating linked data introduces many challenges in both source and destination services as links are severed. We show how Stencil enables correct and efficient data migration between services, how it supports the deployment of new services, and how Stencil could be incrementally deployed.
\end{abstract}



\settopmatter{printfolios=true}

\maketitle

\pagestyle{plain}

\section{Introduction}

The centralization of Internet services and cloud applications has been a boon to users worldwide in recent years. However, with this centralization comes the loss of privacy, the lack of autonomy, and application/service lock-in~\cite{liu2020re}. In recent years, some researchers have argued against centralized approaches, and for decentralized alternatives, but have yet to reach even broad agreement on what should be done~\cite{ali2017blockstack, namecoin, emercoin, mastodon, friendica, identica, riot, matrix, zeronet, hyperboria, cjdns, liu2017barriers, benet2014ipfs, beaker, maidsafe, securescuttlebutt, nextcloud, ring, sia, storj, swarm, filecoin, gnusocial}.  Little progress has been made on the deployment and adoption of decentralized platforms.

There are different flavors of decentralization. Replacing a centralized application or system with a single decentralized alternative is scarcely better: while centralized systems have a host of problems, selection of a single decentralized system is not a viable alternative as it may lead to re-centralization. However, the natural conclusion of this reasoning---a proliferation of decentralized systems, many of which will overlap in their offerings---is both significantly harder for users to navigate and harder for application developers, who have to seemingly build their systems from scratch for each platform.

There are many important technical design challenges in decentralized infrastructure, including security, naming, and more. Our goal is to narrow the focus to the key issues the \emph{architecture} must adjudicate as opposed to an individual application or service.
We argue that we need a \emph{pluralist} architecture: one that allows the co-existence of applications and seamless migration between them. Not only can such an architecture prevent user lock in, but it can also ease the pain of developing decentralized applications. 
Put another way, a pluralist architecture is one that picks no winners: instead, it allows a marketplace of services to be developed, and provides enough scaffolding and restrictions to ensure that the landscape does not become balkanized.

We describe \textbf{Stencil}, a pluralist cloud architecture that enables migration between cloud services such as social media. 
In this context, Stencil addresses the twin problems of a) social media apps making user data hard to download and b) once downloaded, the data is no longer useful because it is no longer linked (to other data held by that user or others). Further compounding this challenge are the network effects that cause new applications to be relatively low value until enough users have bootstrapped the system. For example, Diaspora~\cite{diaspora} has long sought to provide an alternative to Facebook; if Alice wants to try it, she will find when she downloads all her Facebook data that it cannot be easily imported into Diaspora, and even if it were possible, the data would have lost meaning without its links to data of others who did not migrate.

The existence and growth of user lock-in are not entirely due to technical factors. Rather, they stem from the financial incentive for collecting and monetizing user data. 
Changing the prevalent profit or business model is out of the scope of this paper, but policy measures such as the General Data Protection Regulation (GDPR) have forced the hand of large-scale services, ensuring some degree of data portability as a side-effect of privacy objectives \cite{gdpr}. 
Indeed, the providers of some large-scale services are beginning to recognize this same issue, as in 2018, a data transfer initiative was introduced by Google, Facebook, Microsoft, and Twitter, called Data Transfer Project (DTP) \cite{dtp}, in order to enable data migration between various services of these organizations. 
However, DTP does not consider preserving contexts or reconciling semantic differences during and after migration between different services, which could eventually make migrated data useless or cause anomalies. 
Stencil is developed as a solution to the inherent technical problems in service-to-service portability in such applications.

Stencil's pluralism is rooted in the challenge applications face in agreeing upon data formats, and in coordinating any kind of updates or changes. In Stencil, applications are bridged by schema mappings that capture the semantic translation of data from one application to the other. Data can thus be migrated from one application to another through the use of the schema-mapping translation. A naive implementation of migration is likely to produce numerous database- and application-level anomalies, as table rows and data fields are deleted from the source application and inserted into the destination application. Locking all migrating data can prevent transient anomalies of this nature, but not only do locks increase service downtime, but after migration is completed, persistent anomalies must still be handled through additional mechanisms that may not exist natively.

To solve this problem, Stencil requires application writers to be able to specify two things: 1) a directed acyclic graph (DAG) that describes the semantic relationships of data in their own application, and 2) a schema mapping from a source application to their own destination application. Stencil uses the DAG to enforce correctness at both the source and destination applications, and uses the schema mapping to map data from the source to the destination application.

A migration system must protect the security and privacy of user data as data passes through the system. Furthermore, migration or removal of shared data may also lead to a great number of privacy or ownership questions: 1) Can Alice migrate Bob's messages in the same message group? 2) Even if a migration system allows Bob to share his messages to Alice, can Bob still own his messages which has been migrated by Alice to a new application? 3) What if Bob wants to unshare his messages later which have been migrated? Stencil addresses security and privacy challenges using permissions, standard authentication and encryption techniques, and by relying on policy definitions. We defer more subtle issues as a topic for future work; e.g. Bob's posts can only be viewed by Alice in the source application, but the destination application may make Bob's posts public due to the lack of fine-grained visibility.

This paper makes the following contributions. First, we define data migration in a pluralist architecture, identify the inherent technical challenges, and decompose the types of migration-induced anomalies. 
Second, we present the design and implementation of Stencil, which aims to resolve the fundamental challenges in this space, and built a prototype of Stencil in 26,329 lines of Go.  The key, however, is that applications can build upon the Stencil infrastructure and need not make significant changes to their codebases.
Third, we evaluate Stencil by integrating four social network applications with Stencil, each of which only requires a few hundred lines of JSON specification, and demonstrate that Stencil can seamlessly migrate user data without anomalies.

\section{Data Migration} \label{migration}

Here we present three high-level examples of migration, which are not possible today. From these examples, we motivate the definition of three types of migration.
\vspace{1mm}

\noindent 1. Alice uses Facebook but is upset about the Cambridge Analytica Scandal~\cite{cambridge_analytica} and decides to migrate to Twitter. She wants to delete everything including her posts, comments, events, etc., from Facebook, and to automatically migrate as much as possible to Twitter instead of rebuilding her online life all over again. 
\vspace{1mm}
    
\noindent 2. Alice decides to migrate data from Facebook to Twitter since she will mainly be using Twitter. However, she cannot completely leave Facebook because most of her friends and colleagues are still there. She wants to keep everything in Facebook, and to copy as much data as possible from Facebook to Twitter automatically.
\vspace{1mm}

\noindent 3. Alice creates an event in her Facebook, and wants to share the event in Linkedin, Instagram, and Snapchat as well. She wants to keep the event consistent in all applications (e.g., if she changes the event date in Facebook, she wants the change to be applied to the other instances of the event automatically). Furthermore, she also wants users' replies or comments on the event in Facebook to appear in all the other applications.
\vspace{1mm}

Based on these migration use cases, we define three types of migration: \textit{deletion migration}, \textit{independent migration}, and \textit{consistent migration}. Deletion migration entails moving data from source applications to destination applications.  Independent migration entails copying data from source applications to destination applications.
Consistent migration entails copying data from source applications to destination applications, and keeping data in source applications consistent with the copied data in destination applications all the time.
Our current implementation of Stencil handles deletion and independent migrations. Consistent migration requires significant additional support from application developers (e.g., synchronizing data between corporations or unified data stores).

\subsection{Challenges}
In any type of migration, moving data from one service to another faces two main challenges: preserving context and reconciling semantic differences. 
User data is important and valuable for many reasons. The content itself is the creative product generated by each individual while on the platform, the links represent social relationships, and the interactions between users elaborate upon these simple connections to form shared contexts between users. 
Today, large-scale services allow users to download their data; re-uploading content or importing contacts to a new service are also both relatively simple. However, these methods for moving user data break the links between users and the relationships between their data. Thus, the data becomes inert and loses much of its value. A common data migration substrate is needed to preserve the relationships between data as users move between services.

Semantic differences must be negotiated if data is migrated from one application to another. 
\textit{Application semantics} include not only the data models in which an application stores and manages its data but also the rules by which an application implements its services to serve users.
Consider a shared conversation between Alice and Bob. If Alice moves her data to another service, where does the conversation belong? [ownership] Who can still read it? [permissions] Who is Bob on the new service? [identity] If Bob also moves to the new service later, how is the conversation restored? [re-integration] These kinds of issues are greatly eased if there is a common data format or centralized architecture where they could be dealt with by defining additional permissions and rules. However, a pluralist design by definition involves different applications and application semantics. A systematic approach to data migration is necessary to resolve these differences.


\subsection{Anomalies}
\label{migration_problems}

Migrating data between Internet applications can create a wide range of potential anomalies depending on the semantics of the applications as well as on how the migration process is implemented. To explore the problem space, we constructed dozens of examples that would produce migration-related issues and classified them into six broad categories: dangling data, data loss, data ownership, incoherent data, data confusion, and service interruption. Here, we first introduce a simple canonical scenario of each type. From these scenarios, we distill three central design challenges that Stencil or any other system that supports migration must resolve.

In each of these scenarios, Alice wants to migrate her posts and messages from application X to application Y.

\vspace{1mm}
\noindent
\textbf{Scenario 1: Dangling Data.} 
Alice migrates her post from X to Y, but the system does not allow the migration of Eve's comments on Alice's post to Y. Eve's comments become ``dangling data'', which we define as the data that loses semantic meaning because it is missing some other data that it depends on. As a result, X does not display Eve's comments without the corresponding posts and Eve's comments are also useless without the post.

\emph{Observations:} Assuming for the moment that the application semantics on both/either end of the migration make it ok to remove the dangling data, a naive solution to the dangling data problem is to identify and delete dangling data (i.e., garbage collection). 

\vspace{1mm}
\noindent
\textbf{Scenario 2: Data Loss.} 
Alice migrates her post from X to Y, which has attached likes from other users that application semantics allow to be migrated. During migration, these likes may arrive at Y earlier than the post; since Y considers these likes as dangling data, they are deleted.

\emph{Observations:} We could solve this issue perhaps by waiting for some time before deleting the data, but this can cause additional problems as we will see later. Scenarios 1 and 2 suggest that we must consider how two pieces of data are related and how they should be handled before, during, and after migration.

\vspace{1mm}
\noindent
\textbf{Scenario 3: Data Ownership.} 
Alice migrates the whole message history of a message group to Y, including Eve's messages. This scenario is similar to how Facebook allows users to download whole message histories of conversations including messages from others. Eve is surprised to see her messages in Y because she only sent these messages in X.

\emph{Observations:} A naive solution to this problem is that a piece of data can only be migrated by its owner (or creator).

\vspace{1mm}
\noindent
\textbf{Scenario 4: Incoherent Data.} 
In a message group of X, Alice is booking tickets for her friends and asks if anyone can't go. Eve says she can't go. Alice fails to see Eve's reply because she is migrating to Y. If the migration only allows Alice's own data to be migrated, Eve's reply cannot be migrated to Y. Thus, Alice ends up buying too many tickets.

\emph{Observations:} In social media applications, it is essential to read data from different users to understand whole contexts. In this example, ownership is ``correct'', but the context is lost. Scenarios 3 and 4 show the importance of considering how ownership of shared data is handled across applications.

\vspace{1mm}
\noindent
\textbf{Scenario 5: Data Confusion.} 
Alice migrates her long post, with all the comments replying to her post, to Y, but because her post contains a large video, it arrives at Y much later than the corresponding comments. Y allows migrated data to be displayed once data arrives. The comments specific to the ``missing'' post in Y give rise to confusion.

\emph{Observations:} This data confusion is short-term. A naive solution is to lock Alice's data until it arrives in Y.

\vspace{1mm}
\noindent
\textbf{Scenario 6: Service Interruption.} 
Alice migrates her post and corresponding comments. A comment contains a large video, and arrives much later than its preceding comments and the post, which can actually be displayed without any confusion, so there is an unnecessary delay in service.

\emph{Observations:} This service interruption is unnecessary according to the semantics of application Y. As more data is migrated, service interruptions may be more common or longer in duration. Scenarios 5 and 6 illustrate the importance of considering how migration is implemented in relation to application semantics. 

The scenarios highlighted here have been simplified to be illustrative of the problems that can arise during migration, but more pernicious scenarios are possible, particularly in relation to security and privacy. For example, Alice blocks Eve from viewing her posts in X and migrates to Y. Eve wants to exploit the vulnerability of the garbage collection mechanism from Scenario 2, so she waits for some period of time. Y eventually garbage collects the block that prevents Eve from viewing Alice's posts because Y has not seen Eve before, and thus the block is considered dangling. Finally, Eve migrates to Y and is able to see the posts she could not see in X. 

Altogether, these scenarios outline the design space of migration-aware systems. 
Each of the three scenario pairs (scenarios 1 and 2, 3 and 4, and 5 and 6) represents the results of diametrically opposed design decisions that do not explicitly take migration into account. We distill these scenarios into three key migration challenges:

\noindent C1.
\label{mp:dataLossAndSecurityQuestion} How to define, identify, and act upon the relationships between data before, during, and after migration. \emph{(Scenarios 1 and 2)}\\[0.2ex]
\noindent C2.
\label{mp:ownershipQuestion} How to handle the ownership of data in shared contexts. \emph{(Scenarios 3 and 4)}\\[0.2ex]
\noindent C3.
\label{mp:appSementicsQuestion} How to migrate data in the face of application-level semantics while ensuring service uptime. \emph{(Scenarios 5 and 6)}

\section{Stencil Design}
\label{stencildesign}

In this section, we describe Stencil's overall design and how it addresses the migration challenges from the previous section.

\subsection{Explicitizing Data Relationships} 
\label{datadependenciesandmigrationorders}

\begin{figure}[t]
\centering
\includegraphics[width=0.9\columnwidth]{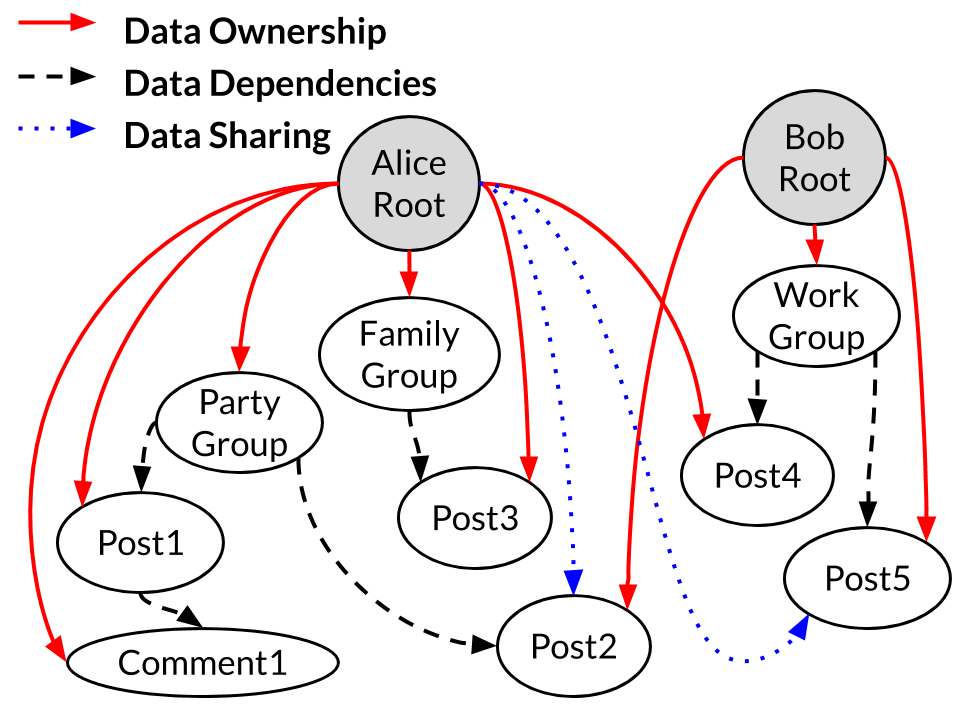}
\caption{DAG of dependencies, ownership, and sharing}
\label{fig:DAG}
\end{figure}

Data in social media applications heavily interlinks with each other. For example, in Facebook, posts may have many comments and comments may have their replies. A reply cannot be displayed without its corresponding comment. We define a \textit{data dependency} as a relationship between data, where one piece of data depends on another to function. In social media applications, replies, comments, and posts are generally authored by specific users; we thus define \textit{data ownership} as a relationship between data and a user, which represents that a piece of data belongs to a user. Finally, we define \textit{data sharing} relationships to be situations where data is shared with a user who is not the owner. 

Data dependencies, ownership, and sharing relationships together form the complete contexts. In other words, these are the relationships between data that are necessary to preserve the semantics of social media applications. Individual pieces of data (e.g., a comment) have very little value in social media applications without corresponding replies, likes, and retweets. Thus, one of our first design decisions is to make explicit these three types of data relationships that are present in social media applications. Once defined, these relationships must then be preserved across application boundaries during and after migration.

These relationships linking data and users form DAGs\footnote{Extending Stencil to general graphs would require atomic migration of cycles.}, and a DAG node, such as \verb=Post1= in Figure \ref{fig:DAG}, is a logical and atomic unit of migration and display.
Application writers need to decide how to group data into nodes based on their application logic because the granularity of grouping data will not only affect correctness but also migration latency; i.e., grouping all data as a node means locking all data before migration, whereas dividing an atomic node into multiple nodes violates atomicity. A root node is a special node containing basic information about users, such as profiles. 
Figure \ref{fig:DAG} illustrates data dependencies; e.g., \verb=Post1= depends on \verb=Party Group=.
Application writers should specify data ownership because it is hard for Stencil to infer or know data ownership by simply examining application-specific attribute names. For instance, Mastodon \cite{mastodonschema} uses \verb=resource_owner_id=, \verb=owner_id=, \verb=account_id=, etc. in different tables to represent data belonging to users, whereas Diaspora \cite{diaspora} uses \verb=author_id= and \verb=user_id=. 

Stencil currently asks application writers to define DAGs in JSON, but we envision semi-automated approaches to streamline this process such as the use of foreign keys to make inferences or static analysis of database queries.

\subsection{Bridging Data Models}
\label{stencilpairwiseschemamapping}

In DTP \cite{dtp}, the project creators envision a structured data model that serves as a common intermediary for migrating data; thus, application writers need to write data adapters to the common data model, and all participating services must agree to adopt this unified data model. This requirement creates a barrier to adoption. 
Rather than a centralized approach taken by DTP, Stencil uses Pairwise Schema Mapping (PSM) \cite{hayek2010data}, which can be defined as follows:

\begin{equation}
a \rightarrow b, b \rightarrow c \Rightarrow a \rightarrow c \end{equation}

\noindent
where \textit{a}, \textit{b}, and \textit{c} represent compatible parts of data schemas in three different applications, and $\,\to\,$ represents transformation. PSM allows an application to transform data into other applications without direct transformation specifications by transforming through compatible data schemas transitively. Since every application writes mappings between a few other similar applications, this path to adoption is decentralized and incremental. 
PSM also reduces the work of defining mappings which can automatically be obtained by streamlining transitive mappings. 

Due to the heterogeneity and complexity of data models, PSM cannot \emph{completely} eliminate the tasks of writing mappings for every other application since it is not guaranteed that each pairwise transformation is lossless. Thus, transitive mappings in PSM might suffer translation errors and intermediate applications cannot act perfectly as a `bridge' to other applications in such cases.
Despite this limitation, PSM allows application developers to select and map to desirable applications with similar data models and thus reduce the more tedious mapping work. Data that cannot be mapped across applications must also be handled properly. In Stencil, this data is stored as a `bag' that the owner can elect to discard, download, and potentially re-integrate later as more compatible mappings become available (Section~\ref{sec:data_bags}).

\subsection{Enforcing Application Semantics} 

So far, we have not distinguished between whether migration induced anomalies occur at the source application or destination application, but anomalies may occur at either or both ends of the migration. Each of the three types of migration (i.e., deletion, independent, and consistent migrations) affects application endpoints differently, but predictably. Deletion migration, for example, has the potential to create anomalies at the source application, but independent migration does not, since data at the source application is untouched.
Our current implementation does not handle consistent migration, but consistent migration is identical to independent migration except that original and migrated data are kept consistent. 
All three types of migration introduce new data in the destination application. Stencil uses two mechanisms to enforce application semantics and optimize for service uptime.

The first mechanism is a \textit{migration order} that ``deletes'' data from a source application during migration and minimizes issues such as data incoherence and service interruption. 
Stencil uses the DAG to determine the migration order as follows: in the source application, Stencil always migrates a piece of data after migrating the data depending on it, and copies the migrating user's root node at first for displaying data in the destination application, but deletes it last.

To preserve destination application semantics, Stencil adds a \textit{migration flag} to the migrated data in the destination service, which exposes the migration process to the destination application. Destination application developers are then free to decide how the system and users can interact with migrated data during migration by annotating the DAG with validation rules. Stencil follows the destination application's DAG, validates migrated data based on the rules defined by application developers, and removes migration flags on valid data to allow it to be accessed by the application. For example, in Figure \ref{fig:DAG}, without posts, comments may not be eligible for display; without the root nodes, all the data owned or shared by the roots may not be displayed.

The order of deleting data from a source application and the order of adding data to a destination application may be in conflict (assuming similar application semantics). For example, a source application wants to delete comments before posts, while a destination application wants to add comments after posts. Thus, if we preserve application semantics, then there is a tension between the source and destination applications with respect to service interruption. Stencil's design preserves service continuation at the source application. Other designs for negotiating this tradeoff are a topic for future work.

\subsection{Migrating Data in Shared Contexts} \label{sec:mig_data_shared_context}

Stencil allows users to migrate data owned (or created) by other users in the same application only if those users explicitly `share' their data with them. For instance, Alice wants to migrate to a new application; her post can be migrated with her as she is the owner, but Bob's comment on her post will only be migrated with Alice if Bob has allowed Alice to do so. 

Migration/deletion of shared data may lead to potential privacy or ownership violations. We describe Stencil's methods of migrating data in shared contexts by discussing four important questions: 1) What shared data rules regarding migration can data owners specify and at what granularity? 2) What happens when different migration types and sharing rules conflict? 3) Will data ownership change after shared data is migrated? 

The granularity of specifying shared data rules depends on applications. Some applications may allow users to choose each specific piece of data that they want to share with other users. Other applications may only allow high-level specifications (e.g., allowing all comments to be shared with the friends group). Stencil provides applications with low-level APIs to store the migration instructions and mark shared data. A piece of data by default has no sharing specifications, and in this case it cannot be migrated by others other than the owner in any type of migration.

While Stencil provides a general way for applications to specify sharing policies, selecting the right policy may not always be straightforward, especially when applications have different semantics. For example, if Alice wants to migrate a post shared with a group of users to another application that does not have group sharing, but only global visibility/invisibility, then this migration could violate the privacy setting of the shared post in the source application. Stencil places the onus of negotiating the differences in application semantics when writing schema mappings on application developers. However, we expect this translation burden to be alleviated as cross-application migration becomes more common and therefore standardized through increased adoption of Stencil.

Sharing data with other users for migration may result in different behaviors in different types of migration. In deletion migration, a comment owned by Bob and shared with Alice will be deleted from the source application if Alice migrates from that application to a new application. In independent and consistent migrations, the comment will stay in both source and destination applications. Stencil allows users to specify the types of migration they want to allow their data to be migrated in. For example, Alice allows her friends to migrate her posts only if they are doing independent migration. If Bob wants to migrate using a different type of migration, Alice's posts will not be migrated. 

After shared data has been migrated to a new application, Stencil enables the original owners of the data to retain their ownership by tracking data relationships before migration, maintaining data transformations, including user and data identity changes, and preserving data relationships after migration. 
Even if data owners have not yet migrated to the new application, Stencil allows applications to use placeholders to re-link data with owners if they join later.
\section{Stencil Architecture}\label{sec:stencil_arch}

\begin{figure}[t]
\centering
\includegraphics[width=.95\columnwidth]{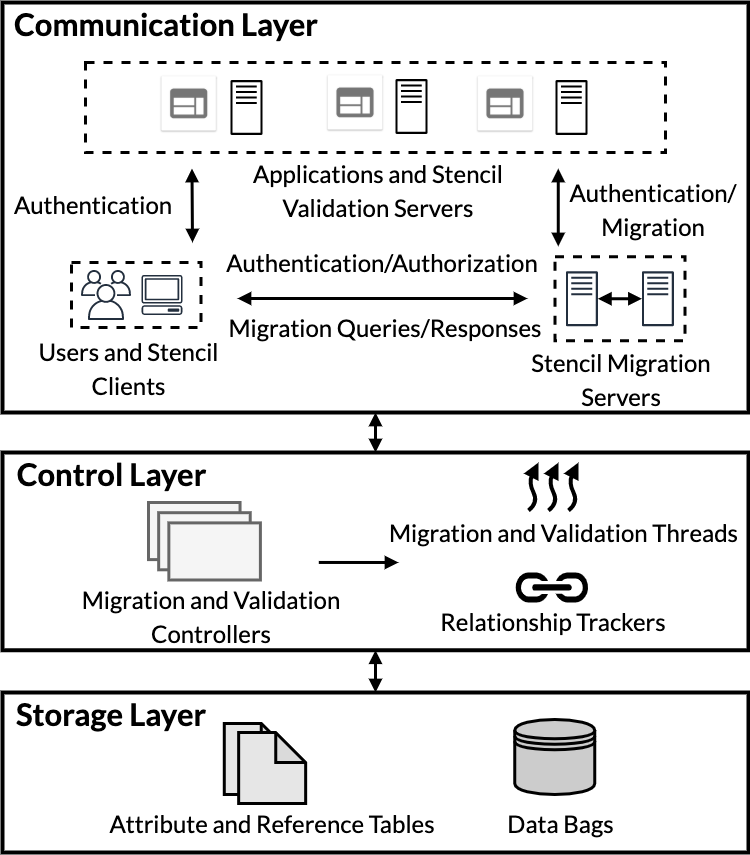}
\caption{Stencil architecture}
\label{fig:stencil_arch}
\end{figure}

Stencil allows applications to use their own data models with data stored in their own storage systems. Stencil uses their data APIs to migrate data between their storage systems. Figure \ref{fig:stencil_arch} shows Stencil's architecture consisting of three layers: a communication layer, a control layer, and a storage layer. 

The communication layer manages the \emph{secure} communication and authentication between different parties. To minimize the barriers to adoption, our current implementation allows users to interact with Stencil through a client running on their local machines. The client authenticates and communicates with one of the Stencil migration servers, and authorizes the servers to perform migrations for users. Since these servers store data relationships and are responsible for migrating user data (although they do not store user data or data bags), for the purpose of data privacy/security, users and applications should only use trusted Stencil migration servers. We expect to bootstrap the overall system by running the initial Stencil migration servers, but we envision these servers to eventually be run by users themselves, trusted third parties, or most likely within the applications themselves as a migration service. Stencil validation servers should be run by destination applications since these servers run a validation algorithm to preserve the semantics of destination applications. 

The control layer consists of migration and validation controllers. 
The controllers run on each Stencil server to create and manage migration and validation threads and relationship trackers locally, and coordinate with other controllers. Based on queries and application specifications, migration threads perform different types of migration, and validation threads validate data to only allow migrated data correctly obeying destination application semantics to be displayed. These threads identify, collect, and re-integrate dangling data.
Relationship trackers are responsible for preserving data relationships after migration. The control layer passes down to the storage layer the data relationships before migration, how data changes during migration, and what data becomes dangling after migration.

The storage layer primarily consists of two tables and data bags. Stencil servers use the reference table and attribute table to store data relationships before migration and data changes during migration respectively. The data in tables is necessary for the control layer to re-link migrated data and re-integrate dangling data. Dangling data is stored in data bags, and can be downloaded and stored by data owners or left in applications according to owner preferences.
\section{Implementation}

\begin{figure*}[ht!]
\centering
\includegraphics[width=1.75\columnwidth]{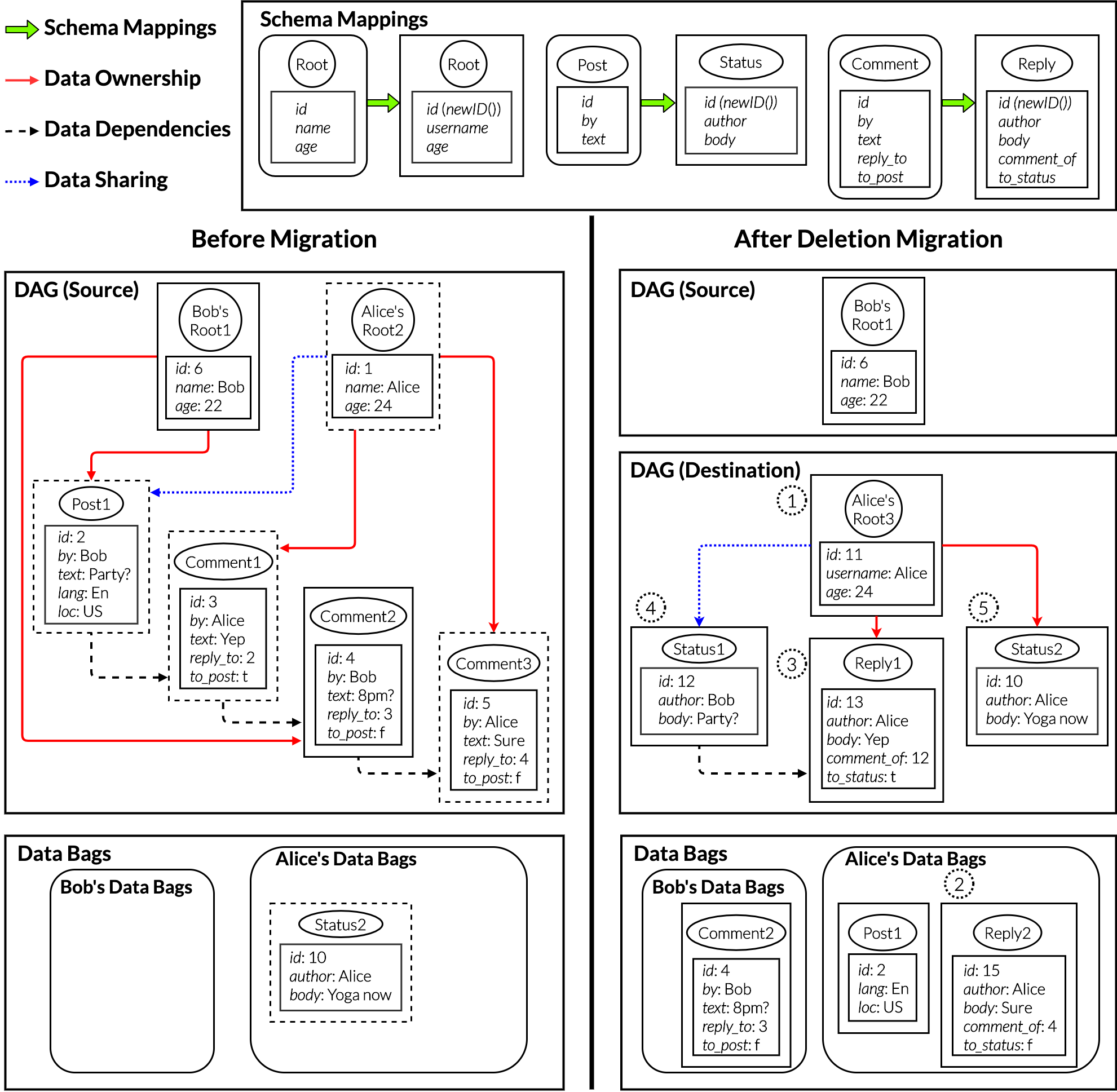}
\caption{A simplified migration example. Alice migrates from the source application to the destination application using Stencil's deletion migration. The Schema Mappings box shows the mappings from the source application to the destination application. Below the Schema Mappings box, the states of the DAGs and Data Bags before and after migration are presented.}
\label{fig:mig_exp}
\vskip -1em
\end{figure*}

We implemented Stencil in 26,329 lines of Go. The implementation consists of several key components: the migration (5,192 lines) and validation controllers (3,797 lines), the relationship tracker (2,070 lines), and PSM (1,381 lines). We use goroutines and channels for migration and validation thread parallelism, and use PostgreSQL in the storage layer to store data. Stencil currently uses a JSON-based format for the specifications of DAGs and schema mappings, and does not handle data model changes. Throughout this section, we will use a running example (Figure \ref{fig:mig_exp}) to explain how migration is implemented in Stencil.

\subsection{Schema Mappings}\label{schema_mapping}

\textit{Schema mappings} define how data can be converted between different schemas in different applications.
The Schema Mappings box in Figure \ref{fig:mig_exp} shows mappings from three nodes in the source application (rounded rectangles) to three nodes in the destination application (rectangles), and unmapped attributes, such as \texttt{Post.loc}, are not shown there. In this example, most data does not change much during migration (e.g., from \texttt{Post.text} to \texttt{Status.body}), except that \texttt{id} of each node has to be re-generated with \texttt{newID()} provided by the destination application to avoid id conflicts. For the attributes, such as \texttt{Reply1.comment\_of}, referring to changed \texttt{id}, Stencil uses a relationship tracker to update them accordingly (Section \ref{sec:relation_tracker}).

\subsection{Migration Algorithms} \label{sec:migration_alg}

Migration controllers can start many migration threads that migrate data concurrently. Each thread runs migration algorithms independently and synchronizes with others implicitly through an application's storage system rather than by sending messages to each other directly, which simplifies the design of algorithms. 
Since Stencil ensures the continuity of application services, migrating all data could take an arbitrarily long time, especially if new data is continually being created. In this situation, the application could allow users to migrate data created before a specified time.
Stencil currently supports deletion and independent migrations, but does not implement consistent migration. In our current architecture, consistent migration would require significant application changes to support data consistency across applications. 

\subsubsection{Deletion Migration}

To preserve the semantics of source application services, a migration thread runs the deletion migration algorithm to 1) move data from the source application DAG in the order in which a node is migrated after migrating the nodes depending on that node, and 2) identify and put dangling data into \emph{data bags}.
During a user's deletion migration, Stencil prevents new concurrent deletion migrations by that user from either the source or destination application to other applications to avoid incomplete migrations due to conflicts.

The example in Figure \ref{fig:mig_exp} shows how Stencil performs deletion migration. In this example, Alice migrates from the source application to the destination application. The DAG (Source) box on the left presents the state of the DAG in the source application before migration, and the dashed rectangles indicate the migrated nodes. The DAG (Destination) box on the right presents the state of the DAG in the destination application after migration. 

A migration thread starts migration by copying \texttt{Alice's Root2} in DAG (Source) on the left to \texttt{Alice's Root3} in DAG (Destination) on the right for validation threads to validate data in the destination application, and then traverses the DAG in DAG (Source) by following the data relationships until it reaches \texttt{Comment3}, which has no next node. The thread migrates \texttt{Comment3} owned by Alice 
to \texttt{Reply2}, but \texttt{Reply2} ends up in Alice's Data Bags on the right. The thread then goes back to either \texttt{Comment2} or \texttt{Alice's Root2} depending on how it traverses the DAG. Supposing it goes back to \texttt{Comment2}, it cannot migrate this node as the node belongs to Bob and is not shared with Alice. Then the thread goes back to \texttt{Comment1}, and finds that it can migrate \texttt{Comment1}. Before migrating \texttt{Comment1}, it needs to put \texttt{Comment2} into Bob's Data Bags on the right since \texttt{Comment2} only depends on \texttt{Comment1} to function and will become dangling after \texttt{Comment1} is migrated. 

Generally, before migrating a node or putting a node into data bags, a migration thread needs to put any other node \textit{only} depending on that node into data bags. 
After migrating \texttt{Comment1} to \texttt{Reply1}, supposing the thread goes back to \texttt{Post1}, it can migrate \texttt{Post1} since the node is shared to Alice. However, there is no mapping for the attributes \texttt{lang} and \texttt{loc} in \texttt{Post1}, so part of \texttt{Post1} is put into Alice's data bags and the other part is migrated to \texttt{Status1}.  (In future work, other than deleting a shared node from the source application, we expect to allow owners to choose to keep a shared node in place if the node is being migrated by others in a deletion migration; however, this will lead to multiple copies of data and thus version control is required.)
Finally, after migrating \texttt{Status2} in Alice's Data Bags (Section \ref{sec:data_bags}), the thread deletes \texttt{Alice's Root2}. 
The DAG (Source) on the right shows that only \texttt{Bob's Root1} is still in the DAG of the source application after migration. 

\subsubsection{Independent Migration}

In independent migration, as data is not deleted in the source application, data will not become dangling due to missing related data. For example, in Figure~\ref{fig:mig_exp}, rather than the DAG (Source) after deletion migration shown in the figure, the DAG (Source) after \emph{independent} migration will be identical to the DAG (Source) before migration. The DAG (Destination) after independent migration will remain identical to the DAG (Destination) after deletion migration. However, after independent migration, the Data Bags will only contain \verb=Post1= and \verb=Reply2=. This is because \verb=Post1= lacks schema mappings for migration to the destination application, and \verb=Reply2= is still dangling at the destination application. Also, Bob's \verb=Comment2= will not be in Bob's Data Bags after migration because it does not become dangling in the DAG (Source) after independent migration. 

There is no need to preserve the migration order in independent migration, or migrate data at a node granularity; instead, to accelerate migration, the independent migration algorithm simply follows data ownership and sharing relationships to copy all data to the destination application.
Given concurrent migration threads, during migration, migrated data in the source application needs to be marked as migrated to avoid being migrated more than once. 
If users migrate back to an application, there could be duplicate data with different identities, which we expect to address in future work using version control.

\subsection{Data Bags}\label{sec:data_bags}

Stencil stores dangling data in \textit{data bags}. Data bags are an abstraction composed of dangling data metadata stored on Stencil servers, and the dangling data itself stored by data owners themselves or left in applications according to owners' preferences. 
Migration threads migrate data not only in the DAG as examined in Section~\ref{sec:migration_alg} but also in data bags. Since migrating data from data bags deletes data in the bags regardless of migration type, Stencil servers only allow a user to perform one migration at a time to avoid conflicts. Migrating data bags has two phases. 

In the first phase, migration threads attempt to migrate data by merging data in data bags and the DAG of the source application to preserve the relationships between data in the two places.
For instance, in Figure~\ref{fig:mig_exp}, if Alice wants to migrate back to the source application later, migration threads will merge \texttt{Post1} in Alice's Data Bags with \texttt{Status1}, migrated from \texttt{Post1} before, and eventually migrate a complete \texttt{Post1} back to the source application.
To achieve this, when data becomes dangling, node IDs are stored with data in data bags. While migrating a node, a migration thread 
tracks data changes (e.g., \texttt{Post1} with \texttt{id} 2 migrated to \texttt{Status1} with \texttt{id} 12), through which the thread identifies, combines, and migrates the related data in data bags and the DAG node. 

In the second phase, after completing the migration of the data in the DAG, migration threads migrate the remaining migratable data in data bags. 
This migration can be in any order and does not influence application services because the remaining data is not related to the DAG. For example, \texttt{Status2} which was put in Alice's data bags in an earlier migration is migrated from Alice's data bags to \texttt{Status2} in the DAG (Destination) in this phase.

\subsection{Validation Algorithm} \label{sec:valid_alg}

As with migration, validation controllers can also start many validation threads running the validation algorithm independently on Stencil servers. To preserve the semantics of the destination application, a validation thread checks and allows migrated data to be displayed in the following order: a node is displayed before the nodes depending on that node (exceptions can be specified in validation settings), and the migrating user's root node is always allowed to be displayed regardless of other nodes. For the data that fails validation, a validation thread puts it into data bags.
To minimize service interruption for users, the validation algorithm is two phases. The first phase is run continuously until the end of a migration to enable a migrating user to use the services of the destination application during migration to shorten service interruption for already migrated data. 
The second phase is run after a migration is completed; validation threads validate the remaining undisplayed data in just one round since there is no need to wait for unmigrated data. 

Following the example in Figure~\ref{fig:mig_exp}, a validation thread randomly picks (more advanced picking rules could be used) and validates migrated but undisplayed data. The numbers in the dotted circles indicate the sequence of nodes arriving at the destination application. Even though \texttt{Reply1} is migrated to the destination application before \texttt{Status1}, a validation thread only allows \texttt{Reply1} to be eligible for display after \texttt{Status1} is allowed to be displayed since \texttt{Reply1} depends on \texttt{Status1} to function. 
According to the validation rules in this example, even though Bob is not in the destination application, \texttt{Status1} is valid for display once it arrives at the destination application because its sharing relationship is satisfied (\texttt{Alice's Root3} arrives first). 
\texttt{Status2} is also valid for display once it arrives because its ownership relationship is satisfied. All the migrated nodes except \texttt{Reply2} can be displayed after passing validation in the first phase; \texttt{Reply2} is eventually put into Alice's data bags in the second phase because the data it depends on is missing.

\subsection{Relationship Tracker}\label{sec:relation_tracker}

Due to the differences of application semantics, migration could break data relationships. For instance, in Figure~\ref{fig:mig_exp}, \texttt{Comment1} is migrated to \verb=Reply1=. If \texttt{Comment1.reply\_to} is directly copied to \texttt{Reply1.comment\_of}, then \texttt{Reply1} will reply to a wrong Status with \texttt{id} 2 instead of \texttt{Status1}. This kind of problem exists not only for data dependencies but also for the ownership and sharing relationships.

Stencil uses a relationship tracker to handle changing identities in these scenarios. To preserve data relationships after migration, the relationship tracker runs during migration and validation to maintain data state changes to re-link data after migration. 

There could be cases where data relating to each other is not in the same application. Stencil allows applications to specify specific placeholders in data attributes to indicate related data is in another application. For example, a comment is not migrated with its corresponding post, so there could be a placeholder in the \texttt{reply\_to} attribute of the comment. 

\subsection{Migration Transaction} \label{sec:migration_transaction}

Since we want Stencil to migrate either all or none, implementing migration as a traditional cross-database transaction would satisfy the requirement. However, this could block reads and writes while migrating bulky data, which interrupts ongoing services and increases user-perceived migration latency. Thus, we implement a user's migration as a \textit{migration transaction}.

In a migration transaction, all migration and display operations are performed in separate and individual database transactions. This ensures the atomicity and durability of migrating each node and also allows users to continue interacting with their data during migration. As we cannot rely on a database to roll back a committed database transaction, Stencil keeps its own write-ahead log to ensure the atomicity of a migration transaction by logging rollback information for each migration operation. To perform a rollback, the migration and validation threads are first stopped. Then Stencil uses the transaction log while traversing the DAG of the destination application to identify migrated data and roll back the migration. Our current implementation does not handle dirty reads at the destination application.
\begin{figure}[t]
\centering
\vskip -0.2cm
\includegraphics[width=\columnwidth]{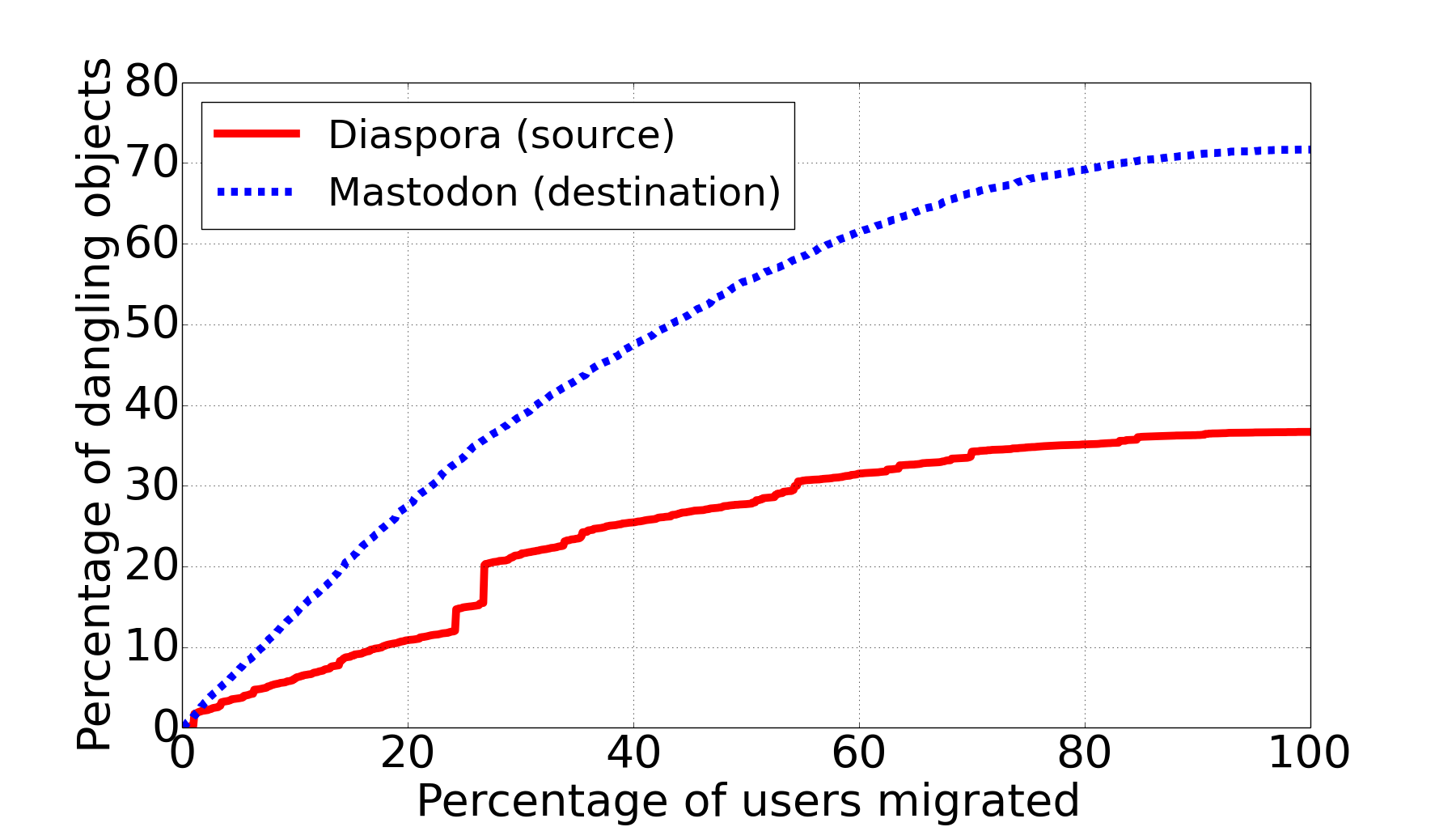}
\vskip -0.2cm
\caption{Anomalies prevented by Stencil in the source (Diaspora) and destination (Mastodon) applications}
\label{fig:anomaliesexp}
\end{figure}

\section{Evaluation}

Since defining scheme mappings and DAGs require us to access application data schemas, we were not able to evaluate Stencil with closed-source applications. Instead, we selected and deployed three open-source applications: Mastodon~\cite{mastodon}, Diaspora~\cite{diaspora}, and GNU Social~\cite{gnusocial}. In addition, we implemented a Twitter clone. We integrated each application with Stencil by writing DAG specifications and schema mappings. 

Since DTP~\cite{dtp} is not yet released, as a baseline for comparison, we implemented a naive migration system without the correctness guarantees and other features of Stencil. The naive system migrates data using the same schema mappings that we use for Stencil. The naive system migrates in arbitrary but intuitive orderings; for example, migration from Diaspora is in the following order: posts, likes, comments, conversations, messages, and users' other data. 

To drive our evaluation, we implemented a social-network data generator in 6,508 lines of Go to generate synthetic data. We generated synthetic datasets based on the prior work analyzing social network data~\cite{leetaru2013mapping, lu2014frequency, wojcik2019sizing}. Each user was assigned a popularity score in the Pareto distribution~\cite{arnold1983pareto}. We then probabilistically assigned posts, likes, comments, etc. to users proportional to their popularity. Conversations pairs were assigned only between friends. 
We generated four Diaspora datasets with 1,000 users (Diaspora 1K dataset), 10,000 users (Diaspora 10K dataset), 100,481 users (Diaspora 100K dataset), and 1,008,108 users (Diaspora 1M dataset). We also generated three datasets each with 10,000 users for the other three applications (Mastodon 10K datasets, GNU Social 10K datasets, and Twitter 10K datasets).
In the Diaspora 1M dataset, there are 2,676,829 follows (81,169 friends), 7,562,681 posts, 30,626,969 likes, 13,481,411 comments, 81,119 conversations, 5,400,995 messages, 46,785,209 notifications, and 3,692,680 photos. In total, the Diaspora 1M dataset has about 51 GB data in the database and 10 TB media on disk. 
Unless otherwise stated, all evaluations used a single migration thread and the Diaspora 1M dataset.

We performed the evaluation in a cloud environment similar to one we expect Stencil would be deployed within. Specifically, we performed migrations in the standalone databases of the test applications, and moved data between databases from a virtual machine (VM) with 128 GB memory and 2 cores to a blade server with 32 GB memory and 16 cores. The two machines were located in a campus data center. 

\subsection{Correctness}\label{eval:correctness}

As the primary focus of Stencil is to address the anomalies considered in Section~\ref{migration_problems}, in our first evaluation, we show what happens when the naive migration system migrated 1,000 users randomly from the Diaspora 1K dataset to Mastodon one at a time. The system deleted dangling data after each migration since it cannot re-integrate dangling data.

Figure~\ref{fig:anomaliesexp} shows the anomalies produced by the naive system through the course of migration until all users are migrated. The total object count in Diaspora is defined as the number of objects before all migrations, and the total object count in Mastodon is the number of objects after all migrations. The percentage of dangling objects is the number of dangling objects produced in an application divided by the corresponding total objects.
We observe that the percentage in Diaspora increases to nearly 40\%. During a user's migration, other users' data may become dangling; e.g., Bob's comments become dangling without Alice's post. Also, some data becomes dangling because there is no mapping from Diaspora to Mastodon. 
The percentage in Mastodon increases to over 70\% because some migrated data fails validation due to lacking the data it depends on; e.g., Alice's migrated likes on Bob's post become dangling because Bob's post is still in Diaspora. 

Stencil prevents all anomalies in both applications by identifying all the dangling data, storing it in data bags, and re-integrating it in future migrations. 

\subsection{Stencil Integration}

\begin{table}[t]
\small
\centering
\caption{Lines of code (LOC) in JSON for DAG specifications}
\resizebox{0.75\columnwidth}{!}{
\begin{tabular}{|c|c|c|c|}
\hline
\textbf{Diaspora} & \textbf{Mastodon} & \textbf{Twitter} & \textbf{GNU Social} \\ \hline
383               & 275               & 254              & 195                 \\ \hline
\end{tabular}
}
\medskip
\label{tab:loc_dependencies}
\end{table}

\begin{table}[t]
\centering
\medskip
\caption{LOC in JSON for schema mappings between applications without PSM}
\resizebox{0.9\columnwidth}{!}{
\begin{tabular}{|l|c|c|c|c|}
\hline
\textbf{}           & \textbf{Diaspora} & \textbf{Mastodon} & \textbf{Twitter} & \textbf{GNU Social} \\ \hline
\textbf{Diaspora}   & -                 & 282               & 197              & 156                 \\ \hline
\textbf{Mastodon}   & 284               & -                 & 162              & 138                 \\ \hline
\textbf{Twitter}    & 228               & 172               & -                & 101                 \\ \hline
\textbf{GNU Social} & 185               & 181               & 107              & -                   \\ \hline
\end{tabular}
}
\label{tab:loc_mappings_w/o_psm}
\end{table}

\begin{table}[t]
\centering
\medskip
\medskip
\caption{
Maximum LOC in JSON for schema mappings between applications with PSM
}
\resizebox{0.9\columnwidth}{!}{
\begin{tabular}{|l|c|c|c|c|}
\hline
\textbf{}           & \textbf{Diaspora} & \textbf{Mastodon} & \textbf{Twitter} & \textbf{GNU Social} \\ \hline
\textbf{Diaspora}   & -                 & 154          & 87        & *                   \\ \hline
\textbf{Mastodon}   & 154          & -                 & *                & 67             \\ \hline
\textbf{Twitter}    & 139          & *                 & -                & 5              \\ \hline
\textbf{GNU Social} & *               & 65             & 22            & -                   \\ \hline
\end{tabular}
}
\label{tab:loc_mappings_with_psm}
\end{table}

A key requirement for the adoption of Stencil is the ease with which cloud services can be integrated with the Stencil architecture. For each of the applications, two of the student authors wrote DAG specifications and schema mappings.
Table~\ref{tab:loc_dependencies} summarizes LOC required to write DAG specifications for test applications.
Each row in Table~\ref{tab:loc_mappings_w/o_psm} shows LOC to directly define schema mappings from the application in that row to the other applications. 

To evaluate how PSM helps ease Stencil integration, we started an experiment by selecting a `new' application to be integrated into Stencil and assumed that the other three applications already had mappings between each other. We then selected a bootstrap application that is the application most similar to the new application since it would make sense to write mappings between applications with similar data models. Among our test applications, Diaspora is most similar to GNU Social, and Mastodon is most similar to Twitter. We then used PSM to derive mappings between the new application and other applications, and calculated the LOC to change derived mappings to be identical to the mappings obtained without PSM. We repeated the experiment four times by treating each test application as the new application. Table~\ref{tab:loc_mappings_with_psm} summarizes the maximum LOC required.
Overall, we find that PSM can reduce the task of specifying schema mappings by roughly half.

\subsection{Dangling Data Reusability}

\begin{figure}[t]
\centering
\includegraphics[width=\columnwidth]{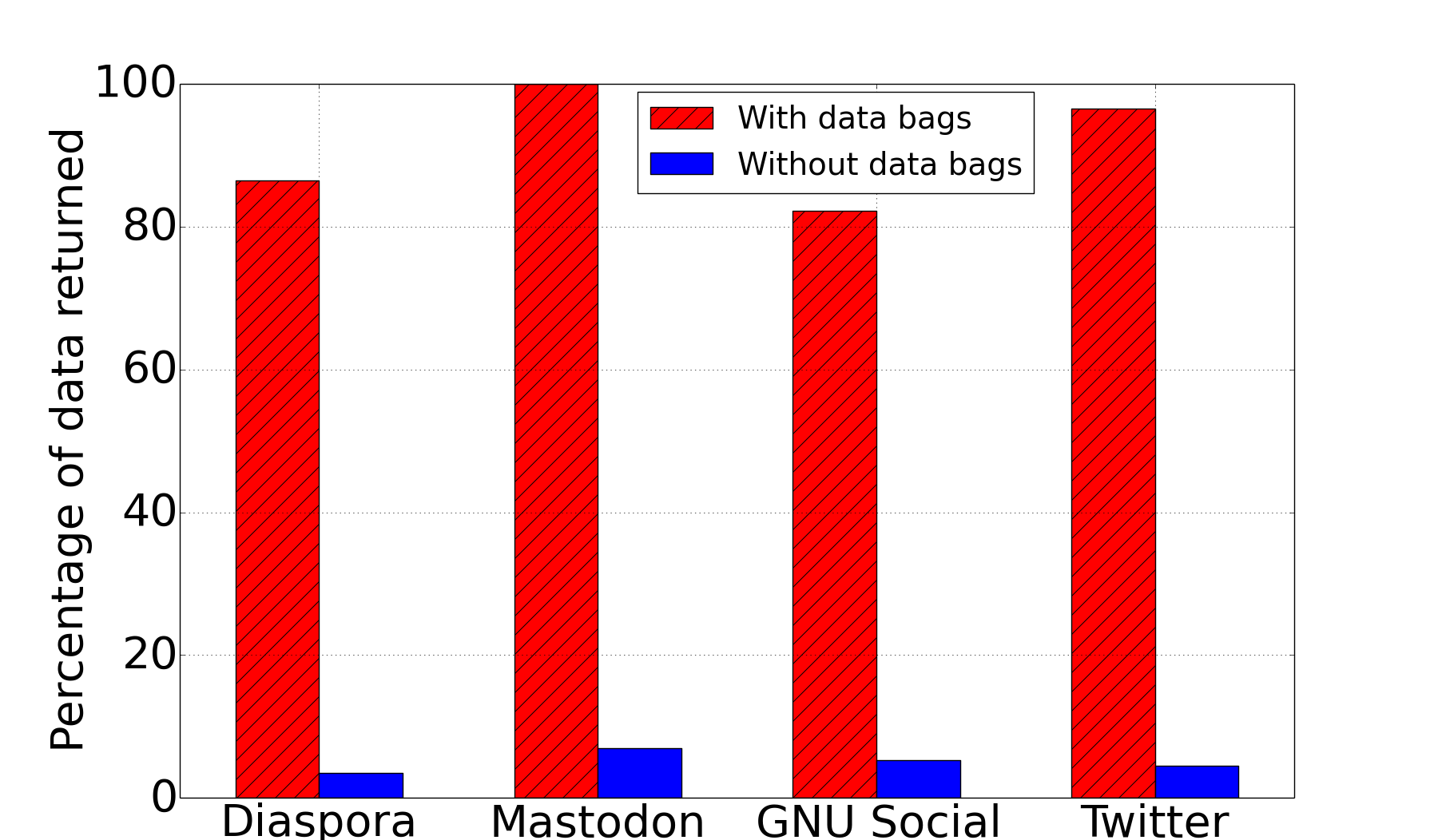}
\caption{
Percentage of data objects originally in each of the four applications that are successfully migrated back to those applications after migrating through a series of three other applications
}
\label{fig:reuse_dangling}
\end{figure}

One of the major benefits of Stencil is the ability to identify, store, and re-integrate dangling data. 
Figure \ref{fig:reuse_dangling} shows how Stencil re-integrated dangling data using data bags after 100 users migrated from each of the four test applications (10K datasets), through a series of the other three other applications, and eventually back to the original application. 
We observe that Stencil with data bags can effectively re-integrate dangling data, and migrate the vast majority of user data back to the original applications. When the starting application is Diaspora, GNU Social, or Twitter, some data cannot be migrated back because our current Stencil implementation does not support Mastodon's DAG specification of owner-less nodes that are shared by multiple users. This causes some nodes to get stuck in Mastodon when data from other applications traverse through Mastodon. We plan to improve Stencil to address this issue in our future work.

\subsection{Migration Rates}

\begin{figure}[t]
\centering
\includegraphics[width=\columnwidth]{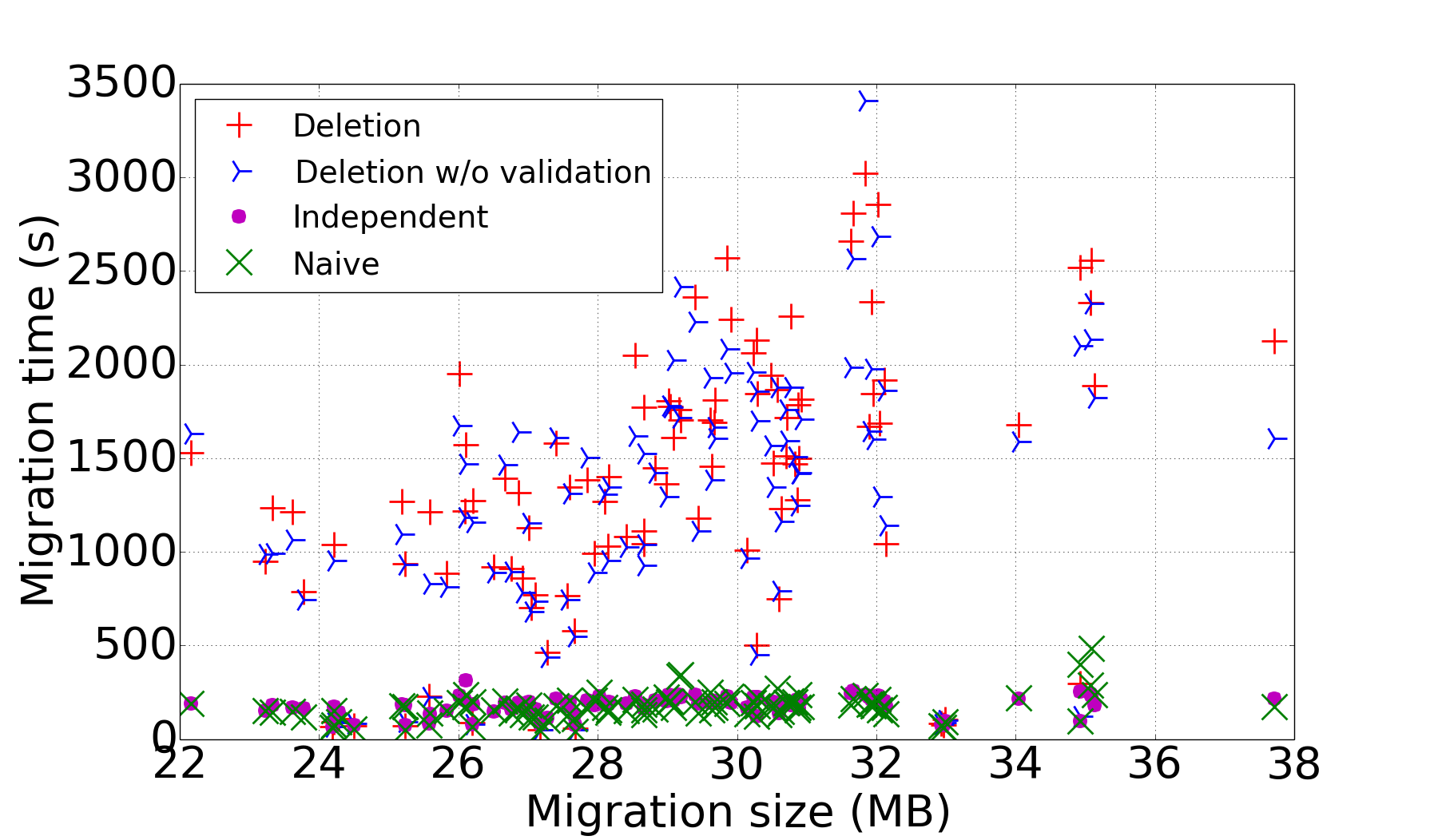}
\caption{Migration rates of different migration types}
\label{fig:migration_rate}
\end{figure}

Figure~\ref{fig:migration_rate} compares migration rates when migrating the same 100 users in the same order using Stencil deletion migrations with and without running the validation algorithm, Stencil independent migrations, and the naive system.
We can observe that running the validation algorithm adds little overhead to migrations but preserves the semantics of the destination application. The independent migrations are about 30 times faster than the deletion migrations mainly because, compared with deletion migrations, independent migrations only need to traverse the data ownership and sharing relationships to copy data. 
The independent migrations can achieve nearly the same performance as the naive system migrations while ensuring correctness.

\subsection{Service Continuity}

\begin{figure}[t]
\centering
\includegraphics[width=\columnwidth]{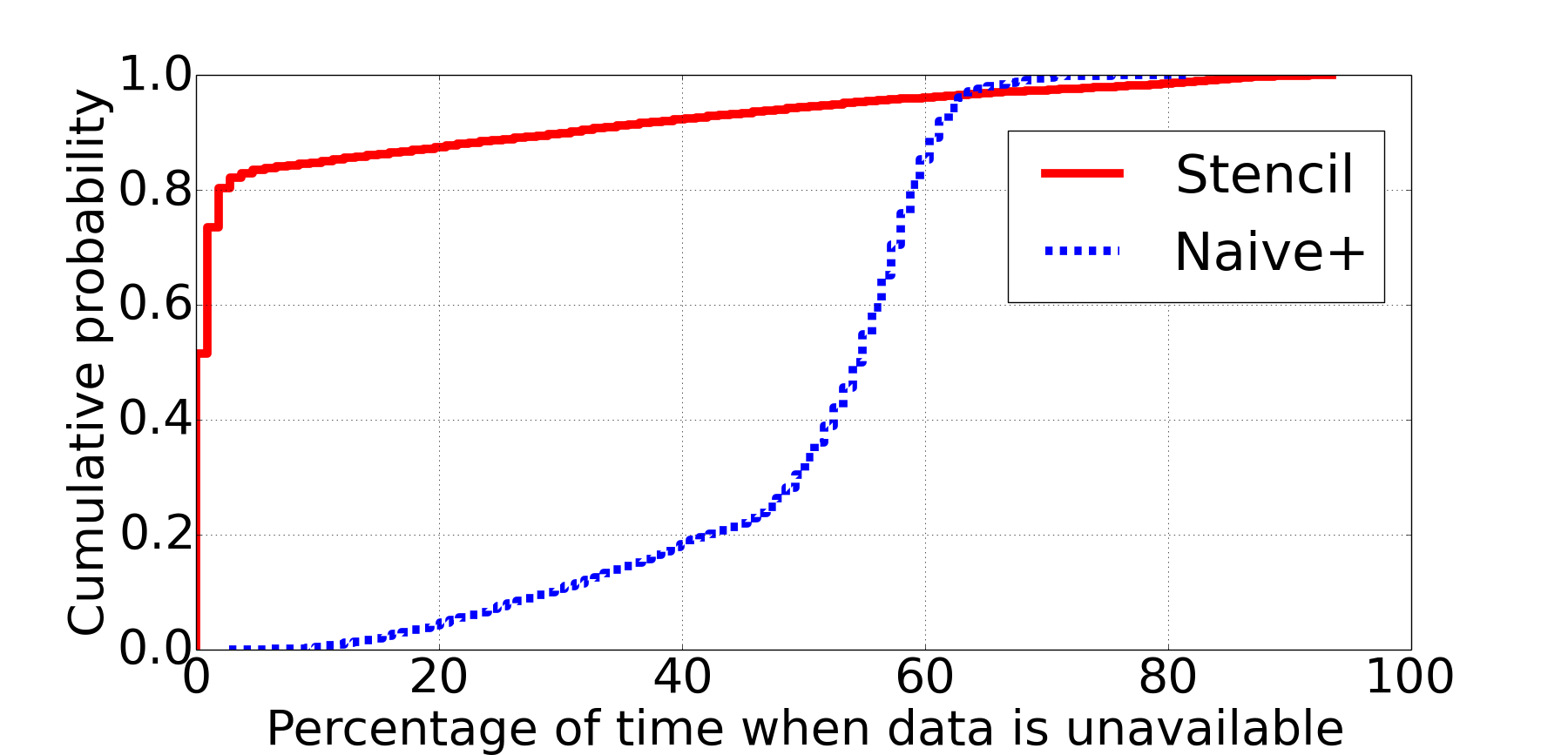}
\caption{Cumulative distributions of the percentage of time when data is unavailable in Stencil compared to Naive+}
\label{fig:service_continuity}
\end{figure}

Figure~\ref{fig:service_continuity} compares how Stencil deletion migration and a naive migration system affect application service continuity.
We define the downtime of a data object as the period of time between when the object becomes inaccessible in the source application, and when it passes validation and is allowed to be displayed in the destination application.
The total time of a migration is the period between when the first object is deleted from the source application, and when the last object is allowed to be displayed in the destination application. 
The percentage of time when data is unavailable is the downtime of a data object during migration divided by the total migration time.
We augmented the naive system by adding a correctness guarantee in the destination application, and call it Naive+. Specifically, we locked up migrated data to prevent the destination application from displaying it until a migration was complete, and then validated and displayed the data using the validation algorithm. We migrated the same 100 users from Diaspora to Mastodon using the two systems to plot Figure~\ref{fig:service_continuity}.

We observe that, compared with Naive+, Stencil effectively reduces the percentage of time when data is unavailable. 
In Stencil, the two-phase validation algorithm continuously allows valid migrated data to be displayed during migration.
There is a long tail, however, since some data cannot be displayed until its related data eventually arrives. 
\color{black}

\subsection{Scalability}

\begin{figure}[t]
\centering
\includegraphics[width=\columnwidth]{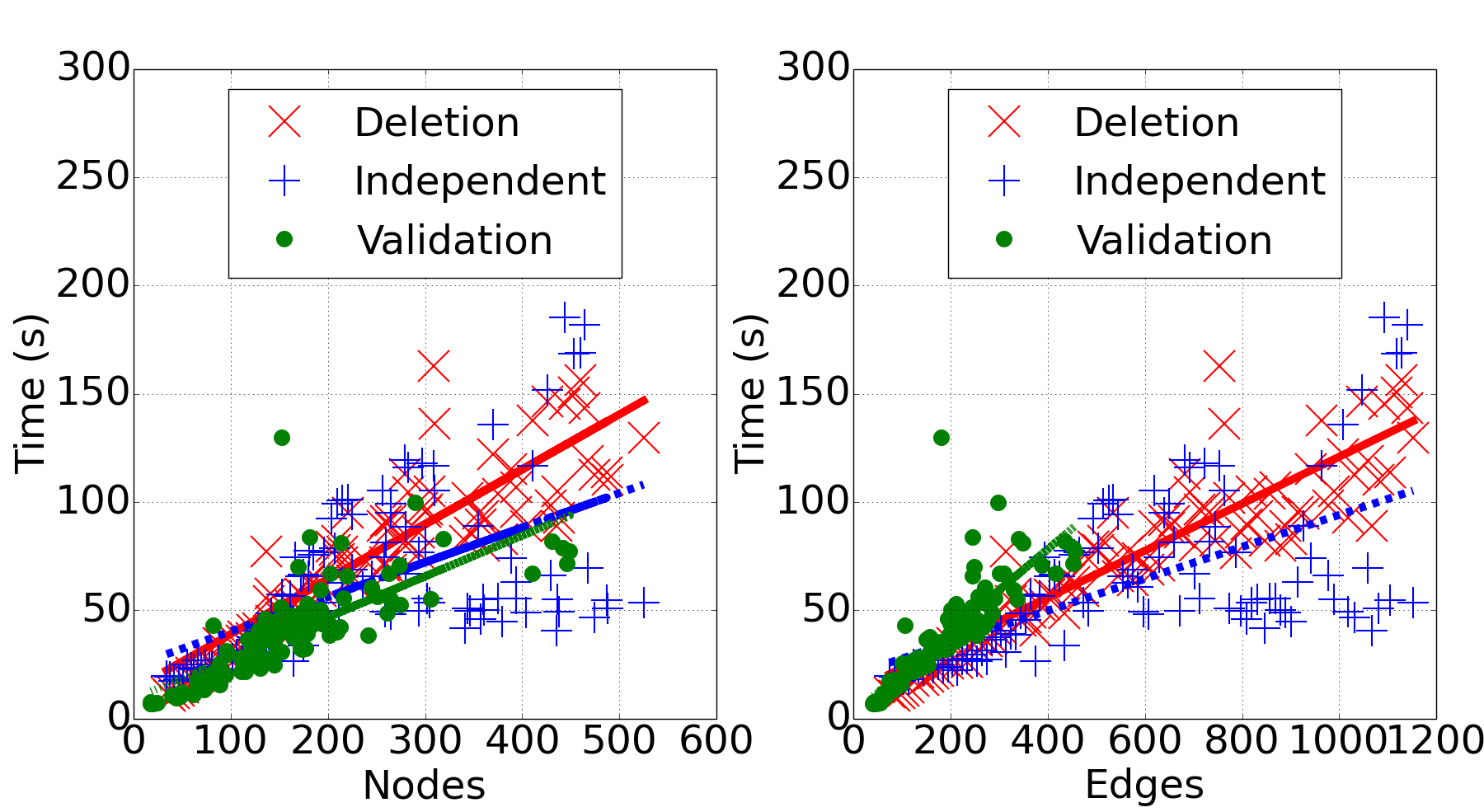}
\caption{Migration and validation times with different numbers of DAG nodes and edges, and each linear regression line fitting the corresponding set of points}
\label{fig:scalability}
\end{figure}

\begin{table}[t]
\vskip 1em
\small
\centering
\caption {Total migration times in the four Diaspora datasets}
\resizebox{0.9\columnwidth}{!}{
\begin{tabular}{|c|c|c|c|}
\hline
\textbf{1K dataset} & \textbf{10K dataset} & \textbf{100K dataset} & \textbf{1M dataset} \\ \hline
22,537.59s               & 17,182.13s               & 22,992.74s              & 73,717.99s        \\ \hline
\end{tabular}
}
\label{tab:datasets}
\end{table}

Figure~\ref{fig:scalability} shows how the deletion and independent migration algorithms and the validation algorithm scale with the number of DAG nodes and edges. We obtained the figure by migrating the same 100 users using Stencil deletion and independent migrations. Before migration, we calculated the number of nodes and edges of a user to be migrated for the migration algorithms. We disabled the validation process until a migration was complete to avoid influencing migration times. When a migration was complete, we calculated the nodes and edges of a migrated user in the destination application for the validation algorithm. A user always has fewer nodes and edges after migration as some data cannot be migrated due to lacking mappings, and a user also has fewer connections with other users in the new application. After that, we started validation threads to validate data there. 
We can see that all the three algorithms scale linearly with edges and nodes. 

To evaluate how the dataset size may influence performance, we selected 100 users with same number of nodes and similar migration sizes across the four Diaspora datasets, and migrated them using deletion migrations. Table~\ref{tab:datasets} shows the total migration times obtained in the four datasets. 
We observe that when the dataset size is below 1 million, it does not have much influence on the total migration time, but when the dataset size reaches 1 million, the total migration time is about 3 times of that in the 100K dataset. This is because basic database operations, such as JOIN and SELECT queries, take much longer as the dataset grows large.
\section{Related Work}

Systems research and practice has cycled between centralized and decentralized designs several times over the past 50 years. 
Recent calls for a transition back to decentralization have often focused on infrastructure---the low-level systems (e.g., distributed ledgers~\cite{wang2019monoxide, lind2019teechain}). 
Some attention has been paid to applications---user-facing code achieving some specific functionality, such as implementing distributed private messaging systems~\cite{Karaoke, tyagi2017stadium, van2015vuvuzela}.  However, not only have these discussions typically occurred in isolation, but also they have been premised on being the \emph{sole} replacement for a prior (current) centralized system.

\subsection{Pluralism in Networked Systems}

The history of popular Web platforms should teach us that picking winners is incredibly difficult. Web search evolved through a bevy of services, from WebCrawler to Altavista to Google.  Social media went through a similar transition, from SixDegrees to Friendster to Myspace to Facebook. Recent fragmentation of the latter---younger users tending to use a variety of services, including Snapchat, Instagram, YouTube, and TikTok---presages the move to pluralism we argue for, but not in the manner one would hope.  
Pluralist architectures have been explored in Internet architecture in the past~\cite{crowcroft2003plutarch, httpnarrowwaist, trotsky}, in which they have been seen as a way to prevent lock-in to any replacement to the IP-based narrow waist. One layer up the stack, pluralism has been explored in Internet routing, with an aim to enable end hosts to select among many possible paths through the Internet rather than a single default path~\cite{detour, deflections, platypus, routebazaar}. However, this concept has yet to be extended to the cloud; in the cloud context, pluralism is harder as applications have greater diversity from the perspective of the underlying architecture.

\subsection{Other Related Work}

Data migration in the context of databases, key-value stores, and similar domains has been studied extensively in various parts of the literature~\cite{pandis2011plp, ousterhout2010case, kulkarni2017rocksteady, mishima2015madeus, elmore2011zephyr, elmore2015squall, das2011albatross, lin2019mgcrab, dtp, clark2005live, liu2009live, kwon2017web}. 
Kown and Moon migrate the states of a web application from one device to another by reconstructing JavaScript closures~\cite{kwon2017web}. 
Rocksteady~\cite{kulkarni2017rocksteady} is a migration technique for the RAMCloud scale-out in-memory key-value store. It aims to migrate data fast while minimizing the impact on response time. Squall~\cite{elmore2015squall} aims to achieve fine-grained reconfiguration in partitioned main-memory DBMSs by interleaving data migration with executing transactions. 
Like Stencil, such work has aimed to handle common problems while migrating data among distributed storage systems, such as live migration (i.e., moving data from the source to the destination while keeping the service alive). However, these systems mainly migrate data within a single application rather than across different applications with different semantics. 
Stencil migrates data between different applications by reconciling semantic differences and preserving contexts.

Tanon et al. designed a tool called Primary Sources~\cite{pellissier2016freebase} to migrate data between collaborative knowledge bases (i.e. Freebase~\cite{freebase} to Wikidata~\cite{wikidata}). Like Stencil, Primary Sources reconciles the differences between two application data models, but their tool also considers the non-technical aspects (e.g. community culture, licensing, and requirements of data). Primary Sources is a crowd-sourced human curation tool to verify data from outside datasets and display it in Wikidata, whereas Stencil is a general architecture designed to migrate data between different data models in ecosystem of social media applications.

The closest related work to ours, DTP \cite{dtp}, attempts to enable users to migrate their data between applications.
Unlike Stencil, however, DTP advocates using a single and standard data model, and requires application writers to write data and authentication adapters translating a given provider's APIs into that standard model. However, such a standard data model would likely lead to limited and inconsistent feature support, and DTP does not consider the anomalies that could arise during migration.

Solid \cite{sambra2016solid} aims to provide data independence from applications. Users store their data in online storage space called personal online datastores (pods). However, pods and applications must be developed by strictly following a number of Solid protocols, which combine several Web standards such as WebID, to enable users to switch between applications. Solid does not directly examine how users' switches between applications affect application services. BSTORE~\cite{chandra2010separating} and Oort~\cite{chajed2016oort} also achieve some degree of data independence from applications by decoupling data storage from applications and by providing unified interfaces for applications to access data, but they don't provide ways for users' data to be reused by applications with different semantics such as data models.

MgCrab~\cite{lin2019mgcrab} uses determinism to maintain the consistency of data on the source and destination nodes. However, it can only be used in a deterministic database system, and cannot be easily applied in the scenarios of data migration between different applications.
Similar to how Stencil uses PSM, MWEAVER~\cite{qian2012sample} applies the concept of PSM in sample-driven schema mappings to generate complete schema mapping paths between tuples through possible pairwise mapping paths between samples.
\section{Conclusion}

In this paper, we have proposed a pluralist architecture that enables the co-existence of social network applications, and seamless data migration between them. We systematically considered a wide range of anomalies that may arise during migration, motivating our design of a migration-aware system. We implemented a prototype of Stencil using a combination of mechanisms to realize migration while handling various possible anomalies. We evaluated Stencil in terms of its correctness as well as ease of integration, performance, and scalability. Based on our evaluation, Stencil can be incrementally deployed for real applications that allows the seamless migration of user data between cloud services.

\bibliographystyle{ACM-Reference-Format}
\bibliography{main.bib}

\end{document}